\documentclass[sn-basic]{sn-jnl}% Basic Springer Nature Reference Style/Chemistry Reference Style
\usepackage{graphicx}%
\usepackage{multirow}%
\usepackage{amsmath,amssymb,amsfonts}%
\usepackage{amsthm}%
\usepackage{mathrsfs}%
\usepackage[title]{appendix}%
\usepackage{xcolor}%
\usepackage{textcomp}%
\usepackage{manyfoot}%
\usepackage{booktabs}%
\usepackage{algorithm}%
\usepackage{algorithmicx}%
\usepackage{algpseudocode}%
\usepackage{listings}%
\theoremstyle{thmstyleone}%
\theoremstyle{thmstyletwo}%
\theoremstyle{thmstylethree}%

\newcommand*\m{\mathrm{m}}                     % metre and milli
\newcommand*\cm{\mathrm{cm}}
\newcommand*\s{\mathrm{s}}                     % second
\newcommand*\kg{\mathrm{kg}}                   % kilogram
\newcommand*\g{\mathrm{g}}
\newcommand*\rmd{\mathrm{d}}                     % d
\newcommand*\rad{\mathrm{rad}}                  % radiant
\newcommand*\Eq{\mathrm{Eq}}                     % Eq
\newcommand*\Hz{\mathrm{Hz}}                   % hertz

\newcommand*\J{\mathrm{J}}                     % joule

\newcommand*\Del{\mathnormal{\Delta}}

\renewcommand{\vec}[1]{\mbox{\boldmath $#1$}}

\usepackage{atbegshi}% http://ctan.org/pkg/atbegshi
\AtBeginDocument{\AtBeginShipoutNext{\AtBeginShipoutDiscard}}

\begin{document}%\addtocounter{page}{-1}
\setcounter{page}{0}
%%\unnumbered% uncomment this for unnumbered level heads

\title[Article Title]{Questions related to the Deflection of Light
by Gravity determined by Soldner, Einstein and Schwarzschild}

%\title[Article Title]{DRAFT: Determining the Deflection of Light without %Einstein´s General Theory of Relativity (GTR)}

%%=============================================================%%
%% GivenName      -> \fnm{Joergen W.}
%% Particle -> \spfx{van der} -> surname prefix
%% FamilyName     -> \sur{Ploeg}
%% Suffix   -> \sfx{IV}

\author*[1]{\fnm{Klaus} \sur{Wilhelm}}
\email{kiwi-nom@t-online.de}.
%{This author mainly considered the aspects of physics and
%the citations in German.}

\author[2]{\fnm{Bhola N.} \sur{Dwivedi}}
\email{bholadwivedi@gmail.com}.
%{This author considered aspects of physics and
%the citations in English.}

\author[3]{\fnm{Karsten} \sur{M\"uller}}
\email{hsk1830@aol.com}.
%{This author mainly considered the mathematical aspects of this work.}

\affil[1]{\orgdiv{Retired from Max-Planck-Institut f\"ur Son\-nen\-sy\-stem\-for\-schung
(MPS)}, \street{Justus-von-Liebig-Weg 3},
\city{37077 G\"ottingen},
\country{Germany}}

\affil[2]{\orgdiv{Retired from Indian Institute of Technology (BHU)}, \city{Varanasi-221005},
\country{India}}

\affil[3]{      %{\orgdiv{....},
\city{Hamburg},
\country{Germany}}

%%=============================================================%%

\abstract{Before we discuss the deflection of light in a gravitational field, we give a brief overview of some basic physical formulas on photon properties, generation and propagation. % in Section~\ref{sec:Introduction}. 
The much debated
problems of the redshift and the photon propagation in a gravitational field is then considered
and applied to the calculation of the speed of light. 
Many citations are given in direct quotations to avoid any misunderstandings. If the quotations are in German, an English translation is provided. 
Based on this speed,
calculated and measured results are recalled on the deflection of light, with emphasis on the deflection near the Sun. We conclude
that the speed of light and the deflection angle can be determined by energy and momentum conservation principles.}

%%=============================================================%%


\maketitle

%%=============================================================%%

\section{Introduction}
\label{sec:Introduction}

The interaction of photons\footnote{\citet{Ein05a} used the expressions ,,Energiequanten''
(energy quanta) and ,,Lichtquant'' (light quantum). The name ``photon'' was
later coined by \citet{Lew26}.} with particles (e.g., atoms, ions or molecules) is governed by equations of the
Special Theory of Relativity (STR) derived by \citet{Ein05a,Ein05b,Ein05c}. 
They can be written as\footnote{See, e.g., \citep{WilFroe}}:
%
%% Equation 1
\begin{equation}
E_\nu = h\,\nu = h\,\frac{c_0}{\lambda}
\label{Photon}
\end{equation}
and
%
%% Equation 2
\begin{equation}
E_0 = m\,c_0^2~,
\label{Energy}
\end{equation}
where $h~=~ 6.626\,070\,15~\times~10^{-34}\,\J\,\Hz$ is the Planck constant, $E_\nu$ is the energy quantum of electromagnetic radiation with a frequency 
$\nu$ and a wavelength~$\lambda$. $E_0$ is the energy of the mass~$m$ at rest. The speed of light in vacuum far from large masses\footnote{Constants are taken from CODATA Recommended Values of the Fundamental Physical Constants: 2022 or from Wikipedia.} is $c_0~=~ 299\,792\,458~ \m/\s$. Energy and momentum of a free massive particle moving with a velocity~$\vec{v}$ relative to a reference frame~S are
\label{Speed_c_0}
%
%% Equation 3
\begin{equation}
E^2 = m^2\,c_0^4 + \vec{p}^{\,2}\,c_0^2
\label{Total_energy}
\end{equation}
with
%
%% Equation 4
\begin{equation}
\vec{p} = \vec{v}\,\frac{E}{c_0^2}~,
\label{Momentum}
\end{equation}
where $E$ is the total energy, $\vec{p}$ the momentum vector ($p = |\vec{p}|$),
and $m$ the ordinary mass, the same as in Newtonian mechanics, cf., ``Letter from Albert Einstein to Lincon Barnett'', 19 June 1948 \citep{Oku89,Oku09}.
With $\beta = v/c_0$, where $v = |\vec{v}| < c_0$, and the Lorentz factor 
$\gamma = (1 - \beta^2)^{-1/2} \ge 1$ it is
%
%% Equation 5
\begin{equation}
E = \gamma\,m\,c_0^2 ~.
\label{Lorentz}
\end{equation}
The kinetic energy of the particle in an inertial system~S is
%
%% Equation 6
\begin{equation}
E_{\rm kin} = E - E_0 = m\,c_0^2\,(\gamma - 1).
\label{kinetic_energy}
\end{equation}

The mass is zero for photons\footnote{A zero mass follows from the STR and a speed of light~$c_0$ in vacuum constant for all
frequencies. Various methods have been used to constrain the photon
mass to $m_\nu < 10^{-49}~\kg$, cf., \citet{Amsetal,GolNie}.}
and Eq.~(\ref{Total_energy}) reduces  to
%
%% Equation 7
\begin{equation}
E_\nu = p_\nu\,c_0~~~~~~~{\rm with}
\label{Photon_energy}
\end{equation}
%%
%
%% Equation 8
\begin{equation}
p_\nu = \frac{h}{\lambda}
\label{Photon_momentum}
\end{equation}
in a region with a gravitational potential~$\Phi_0 = 0$, i.e., far away
from gravitating masses. The gravitational potential is defined as:
%
%% Equation 9
\begin{equation}
\Phi(r) = - \frac{G_{\rm N}\,M}{r}~,
\label{Potential}
\end{equation}
cf., e.g., \citep{Oku00}, with $G_{\rm N} = 6.674\,30 \times 10^{-11}~\m^3\,\kg^{-1}\,\s^{-2}$ Newton's constant of gravity, $M$ the mass of the central gravitating body with a radius\footnote{In line with Eq.~74 on page~{\pageref{Eq_74}}. This $\Delta$ must not be confused with $\Del$ that we use to indicate differences and not the Laplace operator.}~$\Delta$ and $r$ the distance 
from the centre of this spherically symmetric body.
The potential is constraint in the 
weak-field approximation for non-relativistic cases by \\
$0 \le |\Phi(r)| \ll c^2_0$, cf., \citet[Eq.~1,][page~902]{Ein11} and \citet{LanLif}.

To simplify the equations, we set $\Phi_0 = 0$ at $r = \infty$ and write for
$\Phi(r) = \Phi$, unless a specific distance is involved.

%%=============================================================%%

\section{Gravitational redshift}
\label{sec:Gravitation}

The discussion in this section will closely follow our articles on the gravitational redshift \citep{WilDwi14a,WilDwi19,WilDwi20,WilDwi25}.

A relative wavelength increase of $\approx 2 \times 10^{-6}$ was predicted for solar radiation by \citet{Ein08} based on the STR, cf., also \citep{Lau20}.
Experiments on Earth, in space and in the Sun-Earth system have confirmed a relative frequency shift of
%
%% Equation 10
\begin{equation}
\frac{\nu' - \nu_0}{\nu_0} = \frac{\Del \,\nu}{\nu_0}
\approx \frac{\Phi - \Phi_0}{c_0^2}~,
\label{Redshift}
\end{equation}
where $\nu_0$ is the frequency of a certain transition at the gravitational potential~$\Phi_0$ and $\nu'$ is the observed frequency there, if the emission caused by the same transition had occurred at a potential~$\Phi$. 

Before we formulate the physical reasons for the gravitational redshift, several important statements have to be considered:\\
\citet[][p.~892]{Ein05a} pointed out the fundamental difference between mathematics and physics:
\begin{enumerate} \label{Mathematics}
\item[ ] ,,Es ist nun wohl im Auge zu behalten, da{\ss} eine derartige mathematische Beschreibung erst dann einen physikalischen Sinn hat, wenn man sich dar\"uber klar geworden ist, was hier unter ,Zeit' verstanden wird. [...]\footnote{Irrelevant portions of direct quotations are deleted and marked by [...].}.''\\
(One must keep in mind that such a mathematical description makes only physical sense after it is clear what one understands by ``time'', [...].)
\end{enumerate}

In this context, \citet[][p.~111]{Bon86} wrote:  ``A spectral line may be something sounding a little sophisticated, but in fact it is the means of measuring time. Also, whether one is talking of a super-accurate caesium clock, a quartzcontrolled
clock, [...] one is inevitably basing oneself on a source of time affected by the gravitational red shift.''

\citet[][p.~422]{Ein08} wrote with regard to atomic clocks:
\begin{enumerate} 
\item[ ]1. ,,Da der einer Spektrallinie entsprechende Schwingungsvorgang wohl als ein intraatomischer Vorgang zu betrachten ist, dessen Frequenz durch das Ion allein bestimmt ist, so k\"onnen wir ein solches Ion als eine Uhr von bestimmter Frequenzzahl $\nu_0$ ansehen.''\\
(Since the oscillation process corresponding to a spectral line can probably be seen as an intra-atomic process, the frequency of which is determined by the ion alone, we can consider such an ion as a clock with a certain frequency~$\nu_0$.)
\end{enumerate}

We feel that this view merits to be fully appraised, because the electromagnetic forces acting on electrons in atoms or molecules on the Sun are $3 \times 10^{20}$ times larger than the gravitational forces. Nevertheless, \citet[][p.~820]{Ein16a} later concluded that:
%(probably atomic) ``clocks'' would slow down near gravitational centres:
\begin{enumerate}
\item[ ]
2. ,,Die Uhr l\"auft also langsamer, wenn sie in der N\"ahe ponderabler Massen aufgestellt ist. Es folgt daraus, da{\ss} die Spektralinien von der Oberfl\"ache gro{\ss}er Sterne zu uns gelangenden Lichtes nach dem roten Spektralende verschoben erscheinen m\"ussen.''\\
(The clock, therefore, runs more slowly, if it is positioned near heavy masses. Consequently, it follows that spectral lines of light
reaching us from the surface of large stars are displaced towards the red end of the spectrum.)
\end{enumerate}

The first statement is probably correct, if ``corresponding to a spectral line'' is neglected. The second statement is supported by many observations on Earth \citep{PouReb,Craetal,Hayetal,KraLue,PouSni},
in space~\citep{BauWey} and in the Sun-Earth
system~\citep{StJ17,StJ28,BlaRod,Bra63,Sni72,LoP91,Cacetal,TakUen}, if the spectral lines are involved. 

An easy solution to avoid the conflict is to postulate that the oscillating atom, i.e. the `clock', does not necessarily have the same frequency as the emitted spectral line \citep{Bon86,WilDwi14a}. 

Before we study the problem, whether the gravitational redshift is caused by the emission
or the transmission process, the importance of the momentum transfer during absorption or emission of radiation has to be emphasized 
\label{Ein127}
\citep[][pp.~127 and 128]{Ein17}:
\begin{enumerate} \item[ ]
,,Bewirkt ein Strahlenb\"undel, da{\ss} ein von ihm getroffenes Molek\"ul die Energiemenge~$h\,\nu$ in Form von Strahlung durch einen Elementarproze{\ss} auf\-nimmt oder abgibt (Einstrahlung), so wird stets der Impuls~$\frac{\displaystyle {h\,\nu}}{\displaystyle c}$ auf das Molek\"ul \"ubertragen, und zwar bei der Energieaufnahme in der Fortpflan\-zungs\-richtung des B\"undels, bei der Energieabgabe in der entgegengesetzten Richtung.''\\ 
(A beam of light that induces a molecule to absorb or deliver the energy~$h\,\nu$ as radiation by an elementary process (irradiation) will always transfer the 
momentum~$\frac{\displaystyle {h\,\nu}}{\displaystyle c}$ to the molecule, directed in the propagation direction of the beam for energy absorption, and in the opposite direction for energy emission.)
\item[ ]
,,Aber im allgemeinen begn\"ugt man sich mit der Betrachtung des E\,n\,e\,r\,g\,i\,e-Austausches, ohne den I\,m\,p\,u\,l\,s-Austausch
zu ber\"ucksichtigen.''\\ 
(However, in general one is satisfied with the consideration of the e\,n\,e\,r\,g\,y exchange, without taking the m\,o\,m\,e\,n\,t\,u\,m exchange into account.)
\end{enumerate}

The observations of the famous Pound-Rebka experiment \citep{PouReb,PouSni} confirmed in the laboratory the redshift in line with the gravitational potential at different heights.\footnote{In order to reduce the recoil energy to negligible levels, the M\"ossbauer effect was employed \citep{Moe58}. We also want to keep any recoil energy
as small as possible to avoid second order effects and, therefore,
assume $M \gg m \gg \Del m$ in the following calculations.}

\citet[][p.~163]{Craetal} described the result of the experiment as follows:
``From the point of view of a single coordinate system two
atomic systems at different gravitational potentials will have different total
energies. The spacings of their energy levels, both atomic and nuclear, will
be different in proportion to their total energies. The photons are then
regarded as not changing their energy and the expected red shift results only
from the difference in the gravitational potential
energies of the emitting and absorbing systems.''

The same group of experimeters, however, with a different first author,
wrote in a paper by \citet[][p.~165]{Hayetal}:
``In an adjoining paper\footnote{\citet{Craetal}} an experiment is described
in which the change of frequency in a photon passing between two points of
different gravitational potential has been measured.''

This is obviously in conflict with the previous statement, but both conclusions have to be
modified according to a paper by \citet[][p.~2310]{Pou00}:
``My description of our experiment as `weighing photons' is intended to indicate that we could not distinguish between something happening to the propagating photon and to the time scale in the source or absorber. If, in a classical sense, the mass of the photon, as an energy
packet, were falling as a weight, our result would be the same because we do
not independently demonstrate the invariance of the velocity.''

The Pound-Rebka experiment thus quantitatively confirmed the gravitational redshift,
but could not decide whether the shift occurs at the source or on the way to the sensor.
This question was left open by \citet{Dic60}, but
\citet{Oha76} could not find a
loss of oscillations under steady-state conditions
in the Pound\,--\,Repka experiment supporting \citet{Craetal}.
\citet{Okuetal} also concluded that the energy of a propagating photon does not change in a static gravitational field, however, momentum, velocity and wavelength can change. This conclusion is supported by \citet{Qua14} and \citet{Pet01}. 

\citet{Oku00} probably provides the best arguments that the shift occurs at the source:
``The proper explanation of gravitational redshift lies in the behavior of clocks (atoms, nuclei). The rest energy $E_{\rm R}$ of any massive object increases with increase of the distance from a gravitating body because of the increase\footnote{From negative values towards zero; Okun denoted the rest energy by $E_0$ and not by $E_{\rm R}$ ; cf., Eq.~(\ref{Rest_energy}).} of the potential~$\Phi$.''
%
%% Equation 11
\begin{equation}
E_{\rm R} = m\,c^2\,\bigg(1 + \frac{\Phi}{c^2}\bigg) = m\,c^2 + m\,\Phi~.
\label{Rest_energy}
\end{equation}
Okun's rest energy concept corresponds to the statement that a larger amount of energy is available at $\infty$ with $\Phi_0$ than at $r$ with $\Phi$. The energy surplus at $\Phi_0$ obviously is the potential energy, which will be transformed into kinetic energy~$E_{\rm kin}$, see Eq.~(\ref{kinetic_energy}), during the approach to the gravitational centre. If the movement of the mass~$m$ is halted at $\Phi$, the kinetic energy is absorbed by the gravitating mass~$M$ and Eq.~(\ref{Energy}) is still
applicable. This would support Einstein's early assumption that intra-atomic processes would not significantly  be affected by gravity. If photons will be emitted at $\Phi$ with the same energy as at $\Phi_0$, no gravitational redshift would be observed. 

How can this conflict be resolved?

There is general agreement on the fact that the speed of light varies in a gravitational
field, e.g., \citep{Ein11,Sha71,Oku00}. The actual variation is, at this stage, of no importance. 
This speed is often called `Coordinate or World velocity' and is introduced in the following Eq.~(\ref{adjust}) as $c_r$.

We now have to study the photon emission by a massive body~$m$, such as an atom or molecule.
Without gravitational field, i.e. $\Phi_0 = 0$, the rest energy is (the ground state is assumed) $E_{\rm R} = E_0 = m\,c^2_0$. In an excited state the mass
increase is $\Del m$ and the energy~$E'$ of the excited particle is
%
%% Equation 12
\begin{equation}
E' = E_0 + \Del E_0 = (m + \Del m)\,c_0^2~.
\label{excitet}
\end{equation}
A photon emitted at $\Phi_0$ (neglecting any recoil energy) would have the energy 
$h\,\nu~=~p_0\,c_0~=~\Del E_0~=~\Del m\,c_0^2$~. If gravity could not influence the 
emission, the conflict mentioned above would remain. To solve it, we introduced in 
Ref.~\citep{WilDwi14a} the concept of 
an `Interaction region´.\footnote{We follow in this context Einstein's statement
in Nauheim that one could grasp the concept `clock' without detailed information about its mechanism reported in German
by \citet[p. 389,][]{Lau20}} It postulates that
in accordance with Einstein's intra-atomic assumption, the atom tries to emit a photon
with an energy~$p_0\,c_0$ even at $r$ with the gravitational potential~$\Phi$. This is, however, not possible, because the speed of light~$c_r$ at $\Phi$ is smaller than $c_0$  and thus momentum and energy conservation principles
would be violated, cf., page~\pageref{Ein127}. Therefore, an arbitrary differential momentum vector~$\vec{x}$ was introduced to adjust the energy:
%
%% Equation 13
\begin{equation}
||-\vec{p_0}| - \vec{x}|\,c_0 = |\vec{p_0} + \vec{x}|\,c_r ~.
\label{adjust}
\end{equation}
This equation also indicates the momentum adjustment.\footnote{\citet{Fer32} used a similar process to determine the
Doppler shift with the help of energy and momentum conservation.} The interaction region transfers
$-\vec{p_0}$ to the mass~$m$ and can provide +$\vec{p_0}$ for the emitted photon.
In addition, the interaction region transfers $-\vec{x}$ to the mass and +$\vec{x}$ to the photon. This involves an energy tranfer
to the mass of $|-\vec{x}\,c_0|$, which is not available for the photon emission.
This is clearly the cause of the redshift.

Eq.~(\ref{adjust}) allows us to determine $x$ and, in a lengthy calculation, also $c_r$~.
If all the vectors are assumed to be parallel to the emission direction, the equation can be simplified to
%
%% Equation 14
\begin{equation}
(p_0 - x)\,c_0 = (p_0 + x)\,c_r ~.
\label{simplified}
\end{equation}
It is easy to find the solution for $x$ by considering that Eq.~(\ref{simplified}) determines the energy of a photon emitted at $\Phi$. From Eq.~(\ref{Rest_energy}), it follows that the photon energy resulting from $\Del m$ at $r$ is with $\Del m = p_0/c_0$:
%
%% Equation 15
\begin{equation}
h\,\nu' = \Del m\,c_0^2 + \Del m\,\Phi = \Del m\,(c_0^2 + \Phi) = 
p_0\,c_0\,\bigg(1 + \frac{\Phi}{c_0^2}\bigg)~.
\label{Energy_dash}
\end{equation}
Since $h\,\nu'/c_0 = p_0 - x = p_0 + p_0\,\frac{\displaystyle{\Phi}}{\displaystyle{c_0^2}}$,~~~~~~~~~ it is $x = - p_0\,\frac{\displaystyle{\Phi}}{\displaystyle{c_0^2}}$~.

It is now evident, how an atom can sense the gravitational potential: 
The speed of light\footnote{The speed~$c_r$ is
given in Eq.~(\ref{Speed_light}).
Since the very complex process that controles this speed is of no concern in our context, we only refer
to Ref.~\citep{WilDwi19} for an attempt to solve the problem.} $c_r$ in Eq.~(\ref{simplified}) is the answer, because it determines $x$. This solves the dispute between \citet{Mueetal} and \citet{Woletal}, who could not agree on the question, whether or not an atom could sense the gravitational potential.
Wolf et al. correctly assumed that the atom reacted to the potential, but could not present a process to convince M\"uller et al. .
An atom thus reacts to $\Phi$, whereas a pendulum clock depends on
the gravitational acceleration.

The momentum after the emission at $r$ is 
%
%% Equation 16
\begin{equation}
p_r = p_0\,\bigg(1 - \frac{\Phi}{c_0^2}\bigg)
\label{p_Phi}
\end{equation}
and the speed~$c_r$ can be calculated with Eq.~(\ref{simplified}):\\
%
%% Equation 17
\begin{equation}
\frac{c_r}{c_0} = \frac{1 + \Phi/c_0^2}{1 - \Phi/c_0^2} = \frac{1 + y}{1 - y}~,
\label{Calculation_1}
\end{equation}
where we have set $\Phi/c_0^2 = y$. It follows that
%
%% Equation 18
\begin{eqnarray}
\frac{c_r}{c_0} = \frac{1 + y}{1 - y}~\frac{1 + y}{1 + y} = \frac{1 + 2\,y + y^2}{1 - y^2} =
\frac{1 + 2\,y + 2\,y^2 -y^2  }{1 - y^2} = 1 + \frac{2\,y}{1 - y}, 
\label{Calculation_2}
\end{eqnarray}
and, finally,
%
%% Equation 19
\begin{equation}
c_r = c_0\,\bigg(1 + \frac{2\,\Phi}{c_0^2 - \Phi}\bigg)~.
\label{Speed_light}
\end{equation}
The emitted photon energy at the potential~$\Phi(\Delta)$ thus is with $p_{\Delta}$ and 
$c_{\Delta}$:
%
%% Equation 20
\begin{eqnarray}
h\,\nu' = p_{\Delta}\,c_{\Delta} =% \nonumber\\
p_0\,c_0\,\bigg(1-\frac{\Phi(\Delta)}{c_0^2}\bigg)\,\bigg(1+\frac{2\,\Phi(\Delta)}{c_0^2 - \Phi(\Delta)}\bigg) = 
p_0\,c_0\,\bigg(1+\frac{\Phi(\Delta)}{c_0^2}\bigg)
\label{Energy_Delta}
\end{eqnarray}
in agreement with Eq.~(\ref{Energy_dash}). 

This equation could lead to the conclusion that an
energy~$p_0\,c_0$ is obtained for $\Phi(\Delta) \Longrightarrow \Phi_0 = 0$, in contrast to a constant photon energy during the propagation. 

In this context, it is of importance to note that only the speed of light~$c_r$
in Eq.~(\ref{Speed_light}) is controlled by the gravitational field and 
$p_r$ will be adjusted to maintain a constant photon energy. Eq.~(\ref{Energy_Delta}),
therefore, has to be modified to describe the propagation of the photon:
%
%% Equation 21
\begin{eqnarray}
h\,\nu' = p_r\,c_r = 
p_r\,c_0\,\bigg(1+\frac{2\,\Phi}{c_0^2 - \Phi}\bigg) =
p_0\,c_0\,\bigg(1+\frac{\Phi(\Delta)}{c_0^2}\bigg)~.
\label{Energy_r}
\end{eqnarray}
Solving this equation for $p_r$ gives:
%
%% Equation 22
\begin{eqnarray}
p_r = p_0\,\frac{[c_0^2 + \Phi(\Delta)]\,(c_0^2 - \Phi)}
{c_0^2\,(c_0^2 + \Phi)}~.
\label{p_r}
\end{eqnarray}
For a photon emission at $r = \Delta$ this equation is in agreement with Eq.~(\ref{p_Phi}).
A photon emitted at $\Phi = 0$ with an energy~$h\,\nu =p_0\,c_0$ arrives at $\Delta$
with
%
%% Equation 23
\begin{eqnarray}
h\,\nu  = 
p_0\,\frac{c_0^2 - \Phi(\Delta)}{c_0^2 + \Phi(\Delta)}\,c_0\,
\bigg(1+\frac{2\,\Phi(\Delta)}{c_0^2 - \Phi(\Delta)}\bigg) = p_0\,c_0 ~.
\label{Energy_blue}
\end{eqnarray}
and is, compared with the same transition at $\Delta$, blueshifted.

For weak gravitational fields with $|\Phi| \ll c_0^2$, the speed~$c_r$ in
Eq.~(\ref{Speed_light}) agrees with the approximation
%
%% Equation 24
\begin{equation}
c(r) \approx c_0\,\bigg(1 + \frac{2\,\Phi}{c_0^2}\bigg)
\label{Approximation}
\end{equation}
given in many publications for a central gravitational field, e.g., \citep{Oku00,Kraetal,Sha71} and \citet{Sch60}, for a radial propagation\footnote{\citet{Ein12} explicitly stated that the speed at a certain location is not dependent on the direction of the propagation.}. A decrease of the speed of light near the Sun, consistent with
Eq.~(\ref{Approximation}), is also supported by the predicted and
subsequently observed Shapiro delay
\citep{Sha64,Reaetal,Sha71,Kraetal,Baletal,KutZaj}.

An approximation of the
vacuum index of refraction as a function
of the distance~$r$ from a mass~$M$
has been obtained, e.g., by \citet{Booetal,YeandL,Gupetal,WilDwi19},
in agreement with Eq.~(\ref{Approximation}).

%%=============================================================%%

\section{Light deflection near the Sun}
\label{sec:Deflection}

The first quantitative evaluation of the deflection of light
passing close to the Sun was performed in 1801 by \citet{Sol04}.
He wrote:
\begin{enumerate}
\item[ ]
,,Die Kraft, mit welcher der Lichtstrahl [...] angezogen wird,
wird seyn $2{\rm gr}^{-2}$.''\\
(The force attracting the light beam [...] will be $2\,g\,r^{-2}$.)
\end{enumerate}
Soldner obvously assumed an attraction of the light beam by the gravitational field similar to that of a massive body, but it
is unclear, why he wrote 2g, cf., \citet{Tru23,Sau21}. He found a deflection of 
$\omega = 0.84"$ between the emission direction of the beam and the 
the direction at closest approach to the Sun; cf. his Fig.~3.  
The direction observed from Earth is then deflected twice
as much, and would agree with deflection of $1.61" \pm 0.30"$
measured by \citet[][p.~328]{Dys20} during the
total eclipse of May 29, 1919.

\citet{Ein11} also considered the light deflection by gravity.
In his Eq.\footnote{Numbers of equations from different publications are not shown in parentheses.}~3, a speed of light of
\begin{equation}
 3~~~~~~~~~~~~~~~~~~~~~~~~~~~~~~~~c = c_0\,\bigg(1 + \frac{\Phi}{c^2}\bigg)  \nonumber
\end{equation}
is given for a gravitational 
potential~$\Phi$ on page~906. Since $c$ is defined as speed of light
on page~901, the equation should probably read
$c(\Phi) = c\,(1 + \Phi/c^2)$; see also \citet{And17}.
Comparison with Eq.~(\ref{Speed_light}) shows that the second term has only half
the value of $2\,\Phi/c_0^2$.

The difference between $c$ and $c_0$, the constant speed of light introduced on page~\pageref{Speed_c_0},
is described by \citet{Ein11} on page~906 as follows:
\begin{enumerate}
\item[ ]
,,Nennen wir $c_0$ die Lichtgeschwindigkeit im Koordinatenanfangspunkt, so wird daher die Lichtgeschwindigkeit~$c$
in einem Orte vom Gravitationspotential~$\Phi$ durch die Beziehung\footnote{Hier steht Eq.~3, wie oben genannt.}~3
gegeben sein.''\\
(If we call $c_0$ the speed of light at the origin of the coordinate system, the speed of light at a location with gravitational potential~$\Phi$ will, therefore,
be given by the equation\footnote{Eq.~3 shown above. Some confusion can result from the fact that $\Phi$ is positve until the middle of page~904 and
later negative.}~3.)
\end{enumerate}

\citet{Ein11} employed Huygens' principle for his calculation and found: 
\begin{enumerate}
\item[ ]
,,\emph{Ein an der Sonne vorbeigehender Lichtstrahl erlitte demnach
eine Ablenkung vom 
Betrage~$4 \times 10^{-6} = 0,83$~Bogensekunden.}''\footnote{Emphasis by Einstein. The equation is rather strange.}\\
(A beam of light passing near the Sun would be deflected by the amount of
$4 \times 10^{-6}$  corresponding to $0.83"$.)
\end{enumerate}
This result, obtained with a completely different method, is nearly equal to the deflection calculated by Soldner\footnote{It should be noted that Soldner
found for half the deflection $0.84"$. The full deflection would thus agree with the established value, if his factor of 2 in the force equation
could be justified.}. If Einstein had used the light speed of Eq.~(\ref{Approximation}), the result would have been $2 \times 0.83"$ as observed. This follows directly from Equation
\begin{equation}
4~~~~~~~~~~~~~~~~~~~~~~~~~\alpha = - \frac{1}{c^2}\,\int\,\frac{\partial\,\Phi}{\partial\,n'}\,ds
\nonumber
\end{equation}
of the \citet{Ein11} paper, where on pages~906 and 907 he determined the deflection of light in a gravitational field
with the help of H\,u\,y\,g\,e\,n\,s'~principle, if $2\,\Phi$ would be introduced for $\Phi$ .

\citet{Dys20} referred to the paper by \citet{Ein16a}, where on page~822 the result of 1911 is changed to:
\begin{enumerate}
\item[ ]
,,Ein an der Sonne vorbeigehender Lichtstrahl erf\"ahrt demnach
eine Biegung von 1,7" [...].''\\
(A beam of light passing near the Sun suffers a 
deflection of 1.7" [...].)
\end{enumerate}
This value is, within the uncertainty margin, in agreement with the observation in 1919. There is, however, a problem: On the same page, the calculation of the deflection is given by Eq.~74:
\label{Eq_74}
\begin{equation}
B = \frac{2\,\alpha}{\Delta} = \frac{\kappa\,M}{4\,\pi\,\Delta}~.
\nonumber
\end{equation}
$\Delta$ is the distance from a mass~$M$ on page~821 (,, [...] im Abstand~$\Delta$ an einer Masse~$M$ vorbeigeht.''). A more exact definition of $\Delta$ would be: Distance from the
centre of a mass~$M$ (im Abstand~$\Delta$ am Zentrum einer Masse~$M$ vorbeigeht).
$\kappa$ is calculated in Eq.~69 on page~818:
\begin{equation}
\kappa = \frac{8\,\pi\,K}{c^2} = 1,87 \times 10^{-27}
\nonumber
\end{equation}
(shown without unit symbols) and $\alpha$ in Eq.~70a on page~819:
\begin{equation}
\alpha = \frac{\kappa\,M}{8\,\pi}.
\nonumber
\end{equation}
It is to be noted that Einstein did not indicate any unit symbols, but the
value of the constant of gravitation $K = 6,7 \times 10^{-8}$ 
on page~818 can, in comparison with $G_{\rm N}$ in Eq.~(\ref{Potential}), only mean that the cgs system was used. Therefore, 
the solar mass is $M~=~1.988~\times~10^{33}~\g$,
$\kappa~=~1.87~\times~10^{-27}~\cm\,\g^{-1}$, $\alpha~=~1.49~\times~10^{5}\,\cm$ and the 
radius~$\Delta~=~6.957~\times~10^{10}~\cm$. Evaluation of Eq.~74
then yields $B~=~4.26~\times~10^{-6}~\rad$ and thus $0.88"$. This result
is basically the same as that of the 1911 paper.
Where is the mistake?

It might be helpful to note that Einstein obviously used
Huygens' principle also in the 1916 paper as he mentioned on page~821
,,[...] das H\,u\,g\,g\,e\,nssche [sic]\footnote{In direct quotations
typographical and other formal errors are not corrected.
Translations into English, however, avoid formal errors.} Prinzip[...]''.
The same calculations as in the 1911 paper would, together with
the approximation of the speed of light~$c(r)$ in Eq.~(\ref{Approximation}), give 
the correct deflection of $2 \times 0.83"$. However, the speed of light~$c(r)$ 
is not mentioned.

We are not the first to point to this error. \citet{Gin21}, for instance,
wrote in Footnote~2 on page~831:
``It might also be interesting to note that the original Annalen paper by Einstein of 1916 (Einstein 1916, pp. 819–822) has a factor of 2~error in its
eq.~(74), going back to a mistake in its eq.~(70a). This misprint was corrected in the reprint that was included in the collection of papers published as Das Relativit\"atsprinzip, see Collected Papers of Albert Einstein (CPAE), Vol. 6 (Einstein et al. 1996, pp. 334–337) in German and
(Einstein et al. 1997, pp. 196–199) in English.''\\
A question might be, who corrected the misprint and how, since Einstein died in 1955?

The book by \citet{Loretal} could have been helpful, where in the Section: The foundations
of the general theory of relativity by A. Einstein, {\it Translated from ,,Die Grundlage der allgemeinen Relativit\"atstheorie'',
Annalen der Physik, 49  1916}, Eq.~68a on page~159 agrees with Eq.~68a in \citet{Ein16a},
the equations Eq.~69 are also identical, but ``On account of (68a)'' Eq.~70a on page~160 is twice that of Eq.~70a in the 1916 paper.
However, no explanation is given.

Ginoux cited Einstein on page 847:
``Einstein wrote in his conclusion:
`According to this, a ray of light going past the sun undergoes a deflexion of
1.7"...´
Thus, it appears that Einstein's computation of the value of deflection of a light ray
performed in 1915 led him to twice the amount derived in his 1911 paper.
Where does this doubling come from? How did Einstein justify it? [...]
Indeed, as early as 1915, Einstein wrote:
`By use of the Huygens principle, one finds through a simple calculation that a
light ray from the Sun at distance $\Delta$ undergoes an angular deflection of magnitude
$2\,\alpha/\Delta$, while the earlier calculation had given the value $\alpha/\Delta$. 
A corresponding light ray from the surface rim of the Sun should give a deviation of 1.7" (instead of 0.85")´ (Einstein 1915a)''.
Immediately following this correction, \citet{Ein15} continued:
\begin{enumerate}
\item[ ]
,,Hingegen bleibt das Resultat betreffend die Verschiebung der Spektrallinien durch das Gravitationspotential, welches durch Herrn Freundlich
an den Fixsternen der Größenordnung nach bestätigt wurde, unge\"andert bestehen, da dieses nur von $g_{44}$ abhängt.''\\
(However, the result concerning the shift of spectral lines by the gravitational potential, which was quantitatively confirmed by Mr.~Freundlich on stars, will not change, because it depends only from $g_{44}$.)
\end{enumerate}

The speed of light in Eq.~(\ref{Speed_light}) and the approximation in 
Eq.~(\ref{Approximation}) are not deduced from GTR, but from energy and momentum conservation
principles. 
The General Theory of Relativity (GTR) is obviously not necessary
for a determination of the deflection of light near
the Sun.

Page~773 of Einstein's 1916 paper could indicate that it might
not even be possible to get the right answer. Citing E\"otv\"os, it is stated
that in a gravitational field the acceleration is the same for all bodies. This leads to the statement:
\begin{enumerate}
\item[ ]
,,[...] denn man kann ein Gravitationsfeld durch blo{\ss}e
\"Anderung des Koordinatensystems ,erzeugen'.''~\footnote{Bondi (1986) disagreed with this statement.}\\
([...] one is able to `generate' a gravitational field  just by a change of the coordinate system.)
\end{enumerate}

The problem is that \citet{Eotetal} have not considered massless
photons, for which the effect of the gravitational field is obviously twice
as large as for massive bodies. Photons
would behave differently to an accelerated system and a gravitational field; cf., e.g., \citet{Dic60}:``[...] we cannot agree that 
the \emph{equivalence principle} is firmly established by the E\"otv\"os experiment. 
[...], it failed to say anything direct about the propagation of light [...]''. 
\citet{Oha77} also disagreed:
``The strong principle of equivalence is usually formulated as an assertion
that in a sufficiently small, freely falling laboratory the gravitational
fields surrounding the laboratory cannot be detected. We show that this is
false by presenting several simple examples of phenomena which may be used
to detect the gravitational field [...].''
\label{Dic60}

A very precise statement on his
equivalence principle written by
\citet[][pp.~639 und 640]{Ein16b} 
in a reply to
Friedrich Kottler:
\begin{enumerate}
\item[ ] ,,[...]; denn nach meiner Auffassung ruht meine Theorie
ausschlie{\ss}lich auf diesem Prinzip. [...]''.\\
([...]; then in my view my theory is only based on this principle [...].)
\end{enumerate}

\citet{Gin21} goes on to cite Einstein:
``As a result of this theory, we should expect that a ray of light which is passing
close to a heavenly body would be deviated towards the latter. For a ray of light,
which passes the sun at a distance of $\Delta$ sun-radii from its centre, the angle of
deflection ($\alpha$)\footnote{This $\alpha$ is not the $\alpha$ of Eq.~70a.} should amount to
\begin{equation}
\alpha = \frac{1.7\,{\rm seconds\,of\,arc}}{\Delta}~~~\Eq.~23~~~~.''
\nonumber
\end{equation}
This equation is obviously wrong. Not even the unit symbols are in agreement; cf., \citep{BIPM}.

\citet{Wei23} explained on page~11:
``$\kappa$ is a constant that relates the metric curvature to the
stress-energy tensor. It is related to Newton’s gravitational constant and the speed of
light:
\begin{equation}
\kappa = \frac{8\,\pi\,G}{c^4}~~~~~~~~~.''
\nonumber
\end{equation}
It is unclear, where this equation stems from,
but it is inconsistent with Eq.~69 of \citep{Ein16a} given above.\\
On page~14, Weinstein wrote:
``The term $\alpha/r$ in the metric represents the deviation from flatness due to the Sun’s
gravity. $\alpha$ in this term is represented by:
\begin{equation}
\alpha = \frac{\kappa\,M}{4\,\pi} = \frac{2\,G}{c^2}~~~{\rm Eq.~44}.
\nonumber
\end{equation}
Here, $\kappa$ is Einstein’s gravitational constant, $G$ is the Newtonian gravitational constant, $M$ is the mass of the Sun, and $c$ is the speed of light.''\\
This equation is probably correct, but the source is again unclear, because
Eq.~70a of \citep{Ein16a} (shown above) gives half the value of $\alpha$.
Eqs.~257 and 258 on page~61 contain the same error as Ginous' 
Eq.~23.\footnote{Michael D. Godfrey wrote in (eotvos.dm.unipi.it/documents/EinsteinPapers On the In[sic]uence of Gravitation on the Propagation of Light);
``There are two translations of this paper that I know of:
1. [...] Dover Publications, Inc., 1923. This paper appears on pp. 97-108.\\
2. [...] Princeton University Press, 1987.\\
The two English translations appear to have been done independently, although
much of the Princeton text is quite similar to the Dover text. Both just used ´cut and paste´ to produce the Figures. Neither translation seemed to me to be sufficiently accurate to fully convey what Einstein wrote.''. This citation is another indication that there is a lot of confusion in the translated papers of Einstein,}

\citet{Ein20} outlines the effects of the GTR on the light deflection on page~127 in the appendix of his book\footnote{Authorised translation by Robert W. Lawson, D.Sc. (University of Sheffield).}. The first portion of the
statement with the wrong formula of the deflection angle is cited above
by Ginoux. Einstein continued:``It may be added that, according to the theory, half of this deflection is produced by
the Newtonian field of attraction of the sun, and the other half by the geometrical modification (“curvature”) of space caused by the sun.''

Although the book is a translation, we can assume that Einstein checked it and it is difficult to understand that he missed the error in the equation. The sentence added after the equation is, at least, surprising.\\

\citet{EinInf} wrote on page~262:
\label{EinInf}
\begin{enumerate}
\item[ ]
,,Ein Lichtstrahl mu{\ss} im Schwerefeld also genau so von seiner gradlinigen Bahn
abgelenkt werden wie ein K\"orper, der mit Lichtgeschwindigkeit eine waagerechte Bahn
beschreibt.''\\
(A beam of light will be deflected by a gravitational field exactly as a body, which
is moving with the speed of light on a horizontal path.)
\end{enumerate}  
Two remarks may be adequate in this context:\\ 
1. A body moving with the speed~$c_0$ would have
an infinite kinetic energy, which is impossible, and the deflection of the light is probably twice as large; see the next section.\\
2. The second point may be a consequence of
the Soldner and Einstein (1911 and 1916) papers. Soldner considered an attraction
of the light by gravity and, corrected by
the factor~2 error, got the same result
as Einstein 1911 with Huygens' principle
and the wrong speed of light. Twice the
deflection obtained Einstein 1916 [again using 
Huygens' principle obviously with a correct speed of light or its approximation, cf., Eq.~(\ref{Speed_light}) or (\ref{Approximation})],
without, however, giving a formula for the speed.
     
Where did he get this information from? It could be of importance that he received a letter from Karl Schwarzschild on December 22th 1915 with the exact solution of the GTR~equatíons and submitted the report in Berlin on 16th January 1916, cf., \citep{Sch16}.

The above citation by \citet{Gin21} and the quotation from \citet{Ein15}
indicate, however, that twice the deflection was derived 1915 
without any effect on the redshift.

In the book of \citet{Man20}, e.g., we find;
``For light outside the horizon of a BH\footnote{A Schwarzschild Black Hole.} moving radially away from it, we have
\begin{equation}
\frac{\rmd r}{c\,\rmd t}  = \bigg(1 - \frac{2GM}{c^2\,r}\bigg),~~~(60.3)~~[...].''
\nonumber
\end{equation}
This speed of light is obtained with the Schwarzschild metric
and is in agreement with the approximation in Eq.~(\ref{Approximation}).
%The question remains, whether the Schwarzschild equation is also an approximation?

%%=============================================================%%
\section{Discussion and Contradicting Positions}

It is obvious that time presents the main problem area in our discussion. It might be helpful to consider, what 
\citet{Lau59} wrote on the time coordinate added to the three space coordinates:
\begin{enumerate}
\item[ ] ,,[...] MINKOWSKI
(1864 - 1909), der [...] die Zeit als vierte, den drei
Raumkoordinaten gleichberechtigte Koordinate der vierdimensionalen ,,Welt''
einf\"uhrte. Doch handelt es sich dabei nur um einen sehr wertvollen
mathematischen Kunstgriff; Tieferes, wie es manche dahinein legen wollten,
steckt nicht dahinter.''\\
([...] MINKOWSKI (1864 - 1909), who [...] introduced the time as fourth coordinate on an equal footing with the three space coordinates as four dimensional ``World''. However. this was only a valuable mathematical
trick; more, as some want to see in it, is not involved.)
\end{enumerate}
We conclude that in our static system with masses $M$ and $m$ the time coordinate cannot be of major importance, and are convinced that different types of clocks,
such as atomic or pendulum clocks, can react differently to gravitational
forces. The atomic clock will be attracted by the field and has to be
supported by the mass~$M$, but the electromagnetic processes are not 
severely affected. The pendulum clock, on the other hand, is directly
dependent on the gravitational force. The remark by Max von Laue cited by
\citet[][p.~123]{Bon85}:
``I am reminded here of one of the last remarks of von Laue
who at a conference once said 'You must remember, a pendulum clock is not
just the piece that you buy in a shop: it is that plus the Earth'."
is very relevant.

The calculations related to Eq.~(\ref{adjust}) depend on
conservation principles for energy and momentum. We are convinced
that for solar conditions these principles are valid with high accuracy
and leave it open, whether the statement of 
\citet{Str04}
is correct:``[...] general conservation law for energy and
momentum does not exist in GR. This has been disturbing to many people, but
one will simply have to get used to this fact. [...]''.

A comparison between the speed approximation in Eq.~(\ref{Approximation})
and Eq.~(\ref{Speed_light}) shows that the first equations leads to a conflict
for large values of $|\Phi|$, which is not surprizing, but it is remarkable
that the second equation yields $c_r = 0$ for $\Phi = - c_0^2$.

One of the open questions is, why Soldner assumed an attraction of $2\,g$
and obtained the correct deflection\footnote{Soldner's deflection value is often criticized as incorrect in the literature, but it is
the wrong (or not justified) assumption~$2\,g$.
More information on Söldner's evaluation can be found in
\citet{Sau21}.}. Can the following remark of 
\citet[][p.~409]{Car98} 
be an indication that Soldner considered such a ´claim'?\\
``[...] then try to reconcile the results with the
occasional (and not completely unreasonable) ´claim' that objects traveling
at the speed of light fall with twice the acceleration of ordinary matter''.

Dicke's question and Ohanian's statement on page~\pageref{Dic60} about Einstein's equivalence principle (EEP) highlight also our concern.
Photons travelling in an accelerated system without gravitational field horizontal to the accelaration from one side of the lift to the other
will be deflected only half as much as in a gravitational field. It thus seems to be possible to distinguish in the lift between acceleration and gravitation.

The lift experiment can also be performed with photons propagating from the ceiling to the ground; cf., \citep{WilDwi14b}.
If the height is $H$, the photon with an energy~$h\,\nu_0 = p_0\,c_0$ will need a time $t \approx H/c_0$ to reach the ground.
We assume that the lift was at rest in an inertial system during the emission and the gravitational potential is $\Phi_0 = 0$ at the ceiling and 
$\Phi < 0$ at the ground.
The gravitational acceleration then is $- g \approx \Phi/H$.\\ The acceleration of the lift without gravitation is $g$, and
after the time $t$ the bottom reached a speed of $v \approx  g\,t = g\,H/c_0$.

The reflections require some interactions with the walls, and one can compare the temporarily stored energy (converted to mass) or measure with spectrometers the arriving wavelength of the photon and using 
Eq.~(\ref{Photon_momentum}) its momentum.

This constellation is demonstrating two effects:\\
1. If we consider the energies, the gravitational blueshift (cf., Eq.~(\ref{Energy_dash}) for the corresponding redshift)
approximately agrees with
the Doppler effect for small $g$ and $v$ as can be seen from
\begin{equation}
p_{\rm B}\,c_0 = h\,\nu\,{c_0} \approx
h\,\nu_0\,c_0\,\left(1 + \frac{v}{c_0}\right) \approx
p_0\,c_0\,\left(1 + \frac{g\,H}{c^2_0}\right) ~.
\label{n_r_Doppler}
\end{equation}
This might support Einstein's claim.\\
2. However, if the wavelength is measured with a 
spectrometer, one finds with Eq.~(\ref{Photon_momentum})
\begin{equation}
p_{\rm G} \approx p_0\,\left(1 - \frac{2\,\phi}{c^2_0}\right) \approx
p_0\,\left(1 + \frac{2\,g\,H}{c^2_0}\right) ~,
\label{G_Impuls}
\end{equation}
versus $p_B \approx p_0\,(1 + g\,H/c^2_0)$ from Eq.~(\ref{n_r_Doppler}).
This shows that experiments in a lift can determine, whether the lift is accelerated or in a gravitational field.

This can also be deduced from the observation that an energy~$p_0\,c_0$
is transmitted with constant photon energy in a gravitational field,
but $p_B\,c_0 \approx p_0\,c_0\,(1 + g\,H/c^2_0)$ during the acceleration
experiment.

%%=============================================================%%

\section{Conclusion}
\label{sec:Conclusion}

Many errors and misconceptions on the deflection of light by gravity and the related gravitational redshift can be found in the literature, cf., \citep{Lo04}. Two fundamental principles of physics, i.e., the energy and momentum conservations, give, however, a speed of light as a function of the gravitational potential that leads to the observed red- and blueshifts, as well as to a deflection of light\footnote{Employing the calculation of \citet{Ein11} with $2\,\Phi$.} of $1.7"$, which was confirmed during a solar eclipse in 1919 \citep{Dys20}.\\
%%=============================================================%%

\begin{large}
{\bf Acknowledgements}
\end{large}
\vspace{0.5cm}

We want to thank five anonymous skeptic readers who helped to improve the paper, although only some of their proposals were realized by us. We also thank Tilman Sauer for pointing out that we
overlooked in Version~1 his paper:
``Soldner, Einstein, Gravitational Light Deflection and Factors of Two.''

%%%%%%%%%%%%%%%%%%%%%%%%%%%%%%%%%%%%%%%%%%%%%%%%%%%%%%%%%%%%%%%%%
%%%%%%%%%%%%%%%%%%%%%%%%%%%%%%%%%%%%%%%%%%%%%%%%%%%%%%%%%%%%%%%%%

%\bibliography{sn-bibliography}%

\end{document}